\documentclass[aip,jcp,reprint,groupedaddress]{revtex4-1}

\usepackage{graphicx}
\usepackage{bm}
\newcommand{\subscript}[1]{\ensuremath{_{\textrm{\footnotesize{#1}}}}}
\begin{document}

\title{Development of Monte Carlo configuration interaction: natural orbitals and second-order perturbation theory}

\author{J. P. Coe}
\author{M. J. Paterson}%

\affiliation{ 
Institute of Chemical Sciences, School of Engineering and Physical Sciences, Heriot-Watt University, Edinburgh, EH14 4AS, UK.
}%

\begin{abstract}
Approximate natural orbitals are investigated as a way to improve a Monte Carlo configuration interaction (MCCI) calculation.  We introduce a way to approximate the natural orbitals in MCCI and test these and approximate natural orbitals from MP2 and QCISD in MCCI calculations of single-point energies.  The efficiency and accuracy of approximate natural orbitals in MCCI potential curve calculations for the double hydrogen dissociation of water, the dissociation of carbon monoxide and the dissociation of the nitrogen molecule are then considered in comparison with standard MCCI when using full configuration interaction as a benchmark.  We also use the method to produce a potential curve for water in an aug-cc-pVTZ basis.  A new way to quantify the accuracy of a potential curve is put forward that takes into account all of the points and that the curve can be shifted by a constant.  We adapt a second-order perturbation scheme to work with MCCI (MCCIPT2) and improve the efficiency of the removal of duplicate states in the method. MCCIPT2 is tested in the calculation of a potential curve for the dissociation of nitrogen using both Slater determinants and configuration state functions. 
\end{abstract}

\maketitle

\newcommand{\ket}[1]{\left| #1 \right\rangle}

\newcommand{\bra}[1]{\left\langle #1 \right|}

\section{Introduction}

Monte Carlo configuration interaction (MCCI)\cite{MCCIGreer95,MCCIcodeGreer} offers the prospect of capturing many of the aspects of the full configuration interaction (FCI) wavefunction but using only a very small fraction of the configurations.  The method repeatedly performs a configuration interaction calculation with a set of determinants which is enlarged by the addition of random single and double substitutions and reduced by the removal of states which have a coefficient in the wavefunction of magnitude less than a specified cut-off ($c_{\text{min}}$). In principle the method can be applied to ground and excited states of single-reference or multi-reference systems.

Single-point energies have previously been calculated using MCCI,\cite{MCCIGreer95} as have the bond dissociation energies of water and HF.\cite{dissociationGreer}  The errors of electronic excitation energies for atoms computed using MCCI were found to be small when compared with experiment in Ref.~\onlinecite{excite1Greer}.  Electronic excitation energies for small molecules have also been calculated using MCCI with errors of generally circa ten meV when compared with experiment yet using only a tiny fraction of the FCI space.\cite{GreerMCCISpectra}  Potential curves have been calculated for a variety of small systems using MCCI where it was found that non-parallelity errors approaching chemical accuracy when compared with FCI results could be produced using only a very small percentage of the FCI space even when the system was multireference.\cite{MCCIpotentials}  However the calculation of the curve for the very challenging system of fifty hydrogens presented difficulties for the method.  Multipole moments, a non-variational quantity, have also been demonstrated to be calculated satisfactorily by MCCI for ground and excited states using only a very small fraction of the FCI space.\cite{MCCImultipoleandIons} MCCI ionisation energies for atoms have been shown to compare favourably with FCIQMC and exact results, while electron affinities were more challenging with larger percentage errors but the absolute difference was not so poor.\cite{MCCImultipoleandIons}   Again the size of the MCCI space tended to be very small compared to the almost always computationally intractable size of the FCI space.

In this work we consider improving a MCCI calculation by using approximate natural orbitals or second-order perturbation theory.  Although the Hartree-Fock molecular orbitals give the lowest energy single Slater determinant they may not be the most efficient choice for a CI calculation. The natural orbitals (NOs) are the eigenfunctions of the first-order reduced density matrix or one matrix \cite{Lowdin55} which are considered to give better convergence than Hartree-Fock molecular orbitals. One possible benefit is that some natural orbitals may have eigenvalues (occupations) which are essentially zero so can be discarded and hence the size of the FCI space is reduced.  For methods where a wavefunction is not easily available or defined, derivatives may be used to calculate the response or relaxed density matrix which can then be used to give an approximation to the natural orbitals.  It has been demonstrated that the NOs are indeed optimal for a system of two electrons \cite{LowdinShull56} but it is not clear if they are always the best choice for larger numbers of electrons and Ref.~\onlinecite{bytautas:8217} suggests that split-localised orbitals may offer better convergence in larger systems. Approximate natural orbitals have been investigated as a possibly more efficient alternative to variationally optimising orbitals in CASSCF in Ref.~\onlinecite{Abrams04} for a CASCI calculation.  There it was found that potential curves, including the dissociation of ethylene, produced using CASCI with natural orbitals tended to usually have a non-parallelity error of only a few kcal/mol compared with the CASSCF curve.  The exception was the approximate natural orbitals from restricted MP2 which tended to perform poorly while the best results were achieved with CCSD approximate natural orbitals. Natural orbitals from CISD calculation for the ground-state of higher spin states have also been used for excited state MRCI calculations in Ref.~\onlinecite{doi:10.1021/ct200832u}. There excitation energies were found to have a difference of around 0.1 eV when compared with results using CASSCF/MRCI.

We investigate the ability of approximate natural orbitals to improve the efficiency of an MCCI calculation.  We use approximate natural orbitals from Quadratic CI with single and double substitutions (QCISD)\cite{QCISD} or second-order M{\o}ller-Plesset perturbation theory (MP2).\cite{MP2} For a multireference system these methods may perform poorly or even fail to give sensible results with regards to the energy so we also consider approximate natural orbitals from an MCCI calculation.
As a fairer comparison than just a single energy calculation, we consider if MCCI with natural orbitals can offer improvements in accuracy and calculation time compared with standard MCCI for potential curves of water, carbon monoxide and the nitrogen molecule when using FCI as a benchmark.

We saw in Ref.~\onlinecite{MCCIpotentials} that potential curves for small systems for which full configuration interaction results were available could generally be calculated to relatively high accuracy using MCCI.  This was achieved with a very small fraction of the states and it is interesting to consider whether results can be improved by using the MCCI wavefunction as the starting point for a second-order perturbation calculation. To this end we adapt a second-order multireference perturbation method\cite{HarrisonFCIperturbation} to work with MCCI and improve the efficiency of the removal of duplicate states.  This method estimates the energy contribution from the neglected states in a MCCI wavefunction at the expense of the final energy not being variational nor being easily associated with a wavefunction.  We test the assumption that the MCCI wavefunction will be a very good starting point for this perturbation so that MCCIPT2 should be able to produce a more accurate potential curve than MCCI, when both are compared with FCI, by accounting for more of the neglected dynamic correlation from a MCCI calculation.
\section{Methods}

\subsection{MCCI}

The algorithm\cite{MCCIGreer95,MCCIcodeGreer} for MCCI is that the current MCCI wavefunction (usually initially comprising the occupied Hartree-Fock orbitals) has configuration state functions (CSFs) consisting of random single and double substitutions added to it.  These substitutions are definitely attempted in CSFs with a coefficient of magnitude greater than a certain value while other CSFs have a $50\%$ chance of a substitution occurring.  The Hamiltonian matrix and overlap matrix are then constructed and the new wavefunction is found.  Newly added states whose absolute value of coefficient is less than $c_{\text{min}}$ are discarded and the process continues.  Every ten iterations all states are considered for removal, not just newly added ones.  This also occurs on the second last step and no states are added or removed on the final iteration.

In this work we also consider a version of MCCI with a modified behaviour for the removal/addition of states, a convergence criterion and we also use Slater determinants (DETs) instead of CSFs for some computations, including the calculation of the MCCI NOs.  When using DETs the MCCI wavefunction is not necessarily an eigenfunction of $\hat{S}^{2}$ and more states may be required but the construction of the Hamiltonian matrix and first-order reduced density matrix is much simpler.  We calculate the Hartree-Fock molecular orbital integrals using MOLPRO.\cite{MOLPRO}  In this work the initial MCCI wavefunction is the CSF or DET formed from the occupied restricted Hartree-Fock orbitals.
\subsection{Natural orbitals}

The first-order reduced density matrix or one matrix is defined as
\begin{eqnarray}
\nonumber \gamma (\vec{x}_{A},\vec{x}_{B})=N\int \Psi^{*}(\vec{x}_{A},\vec{x}_{2},\cdots,\vec{x}_{N}) \\
\Psi(\vec{x}_{B},\vec{x}_{2},\cdots,\vec{x}_{N})d\vec{x}_{2}\cdots\vec{x}_{N}
\end{eqnarray}
which can be written in terms of the $M$ one-particle molecular orbitals as
\begin{equation}
\gamma (\vec{x}_{A},\vec{x}_{B})=\sum_{i=1}^{M}\sum_{j=1}^{M}\phi_{i}^{*}(\vec{x}_{A})\bm{\gamma}_{ij}\phi_{j}(\vec{x}_B).
\end{equation}
  We use MCCI with Slater determinants and a not too onerous number of iterations with the same $c_{\text{min}}$ as the full MCCI calculation to construct approximate natural orbitals.   We create the one matrix in the one-particle representation using the following method, beginning with $\bm{\gamma}=0$. We consider all the DETs forming the wavefunction. DETs $i$ and $j$ in maximum coincidence only contribute to $\bm{\gamma}$ if they either have no differences to give
\begin{equation}
\bm{\gamma}_{mm}\rightarrow\bm{\gamma}_{mm}+e_{p}c_{i}^{*}c_{j},
\end{equation}
where $m$ runs over all orbitals in the DET, or one difference due to orbitals $k$ and $l$ which results in 
\begin{equation}
\bm{\gamma}_{kl}\rightarrow\bm{\gamma}_{kl}+e_{p}c_{i}^{*}c_{j}.
\end{equation}
Here $e_{p}$ is the sign due to putting the Slater determinants in maximum coincidence.  We average over spins and then diagonalise the one matrix to give the MCCI natural orbitals.  As we cannot be sure that a very small occupation is due to the approximation rather than being something that would occur in the FCI natural orbitals, then we include all the approximate natural orbitals. We recalculate the one and two-electron integrals using these approximate natural orbitals in MOLPRO\cite{MOLPRO} then use them in a longer MCCI calculation using either DETs or CSFs.

We also consider the approximate natural orbitals from QCISD and MP2 calculations. QCISD can perhaps be thought of as a less complex approximation to coupled cluster singles and doubles (CCSD).\cite{CCSD}  In QCISD size consistency is introduced into a configuration interaction method but at the expense of the energy no longer being variational. MP2 uses the Hartree-Fock Hamiltonian as the zeroth-order approximation in a second-order perturbation to give an efficient way to account for some of the correlation.  It is also size consistent but not variational.
Using MP2 or QCISD the one matrix may be approximated by the response one matrix to give approximate natural orbitals which we generate with MOLPRO.\cite{MOLPRO}  Although the natural orbitals we consider are approximate we shall refer to them as the natural orbitals from a certain method, e.g., MCCI natural orbitals.

\section{Single-point calculation using Natural orbitals}
\begin{figure}[ht]\centering
\includegraphics[width=.45\textwidth]{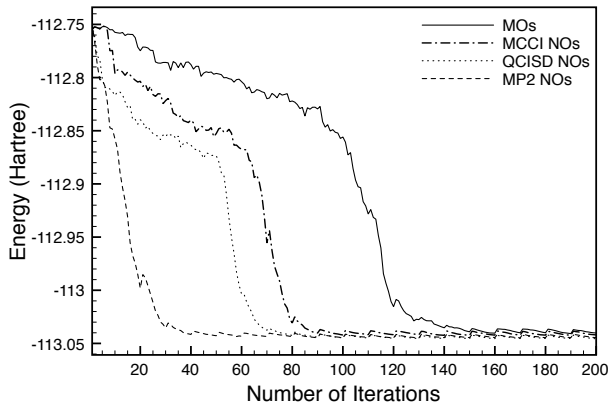}
\caption{MCCI ($c_{\text{min}}=5\times10^{-4}$) energy against number of iterations for CO at $R=2.1316$ Bohr with a cc-pVDZ basis set and with two frozen core orbitals when using either MOs, MP2 NOs, QCISD NOs or MCCI NOs.}\label{fig:NatorbCOiterations}
\end{figure}

We first consider carbon monoxide at its experimental equilibrium geometry\cite{COdipoleExperiment} with a cc-pVDZ basis, $c_{\text{min}}=5\times10^{-4}$, and the two lowest energy MOs or two most occupied NOs frozen. We use 500 iterations of MCCI with Slater determinants, but we see in Fig.~\ref{fig:NatorbCOiterations} that the calculations have essentially converged in much fewer iterations. The MCCI natural orbitals are calculated using a fifty iteration MCCI run.

There is substantially faster convergence per iteration when using approximate natural orbitals here as can be seen in Fig.~\ref{fig:NatorbCOiterations}.  It appears that MP2 natural orbitals followed by those of QCISD  and then MCCI all offer superior convergence to MOs here.  However neither the time cost per iteration nor the overhead from calculating the natural orbitals is taken into account.  One approach is to consider, to three decimal places, the highest final iteration energy and check how long it takes a calculation to first reach this energy or lower on the step after all states have been considered for removal.  Here the final lowest energy ($-113.042$ Hartree) was from using QCISD NOs followed by MP2 NOs while the highest was for MOs ($-113.036$ Hartree). In Table \ref{tbl:COeqTime} we display the time and number of DETs required to reach this energy.  The time for the initial creation of the one and two electron integrals is not included as it is the same for each method.  For MCCI NOs the integrals need to be recalculated and the time cost of this is included as is the QCISD and MP2 calculation time. We see that MP2 NOs took the least time while QCISD NOs used the fewest number of DETs.  MCCI NOs were an improvement of the time and number of DETs compared with MOs but performed less well than the other two types of approximate natural orbitals we considered. As a system moves away from equilibrium it may be that the most efficient NOs are not from the same method.  

\begin{table}[h]
\centering
\caption{Total time and number of DETs for CO to reach $E<-113.036$ Hartree on the step following when all states have been considered for removal.} \label{tbl:COeqTime}
\begin{tabular*}{8.5cm}{@{\extracolsep{\fill}}lcc}
\hline
\hline
Orbitals & DETs & Time (seconds)  \\
\hline
  MOs  & 9,919  & 248  \\
  MCCI NOs & 9,019 & 173  \\
  MP2 NOs & 7,453 & 115 \\
  QCISD NOs & 7,272 &  130 \\
\hline
\hline
\end{tabular*}
\end{table}

\begin{figure}[ht]\centering 
\includegraphics[width=.45\textwidth]{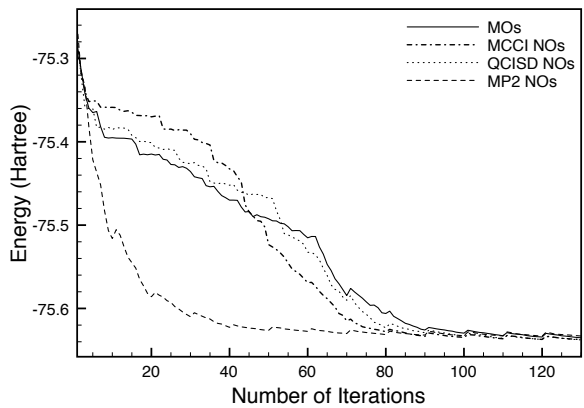}
\caption{MCCI ($c_{\text{min}}=5\times10^{-4}$) energy against number of iterations for C\subscript{2} at $R=1.6$ angstroms with a 6-31G* basis set when using either MOs, MP2 NOs, QCISD NOs or MCCI NOs.}\label{fig:NatorbC2iterations}
\end{figure}

We now consider the carbon dimer with no frozen orbitals and a bond length of $R=1.6$ angstroms.  Here the system is moving away from the equilibrium geometry of $R\approx 1.25$ angstroms. We use the 6-31G* basis set and $200$ iterations with a cut-off value of $5\times10^{-4}$.  In this system we see in Fig.~\ref{fig:NatorbC2iterations} that the fastest improvement per iteration is when using MP2 NOs, then it appears that MCCI NOs followed by QCISD NOs improve on the convergence per iteration compared with MOs.  However we find that the MP2 natural orbitals actually give the highest energy on the final step of $-75.628$ Hartree.  

We see in Table \ref{tbl:C2eqTime} that now MP2 NOs produce the longest time to reach the specified energy but they still use fewer DETs than the canonical molecular orbitals.  MCCI NOs now perform the best, with regards to time and   number of states to reach this energy, but there is not much difference between the results using MCCI NOs and those using QCISD NOs.

\begin{table}[h]
\centering
\caption{Total time and number of DETs for C\subscript{2} to first reach  $E<-75.628$ Hartree on a step following the consideration of all states for removal.} \label{tbl:C2eqTime}
\begin{tabular*}{8.5cm}{@{\extracolsep{\fill}}lcc}
\hline
\hline
Orbitals & DETs & Time (seconds)  \\
\hline
  MOs  & 15,045  & 628  \\
  MCCI NOs & 11,755 & 377  \\
  MP2 NOs & 13,795 & 1134 \\
  QCISD NOs & 11,953 &  381 \\
\hline
\hline
\end{tabular*}
\end{table}

We note that if we pick a high enough energy then MP2 natural orbitals may give the fastest convergence and, in addition, for a single calculation the stochastic nature of the algorithm could affect the order of the methods.  So for  a fairer comparison we will now consider potential curves.  As this requires numerous single-point calculation then any random improvements or deteriorations in the speed of the calculation should average out and rather than considering an arbitrary energy as the target we will use a convergence criterion for each single-point calculation.  In addition to possible faster calculations and fewer states this should enable us to see what, if any, improvement the use of natural orbitals produce in the accuracy of the potential curves.

\section{Potential energy curve comparison}

We now use CSFs in the main MCCI calculation, but we still approximate the MCCI natural orbitals using a DET MCCI calculation. We introduce a convergence check in that the calculation for each point is run until the maximum difference in the last three energies following steps where all states are considered for deletion is $10^{-3}$ Hartree.  Furthermore the MCCI method here is such that no new states are added on any iteration following a step where all states have been considered for removal; previously this only occurred on the last iteration.  This ensures that only the energies of wavefunctions where all states satisfy the cut-off requirement are being compared.  We use twelve processors for the MCCI calculations except for the construction and diagonalization of the one matrix which is carried out in serial.

We quantify the accuracy of the potential curves when FCI results are available using the non-parallelity error (NPE)\cite{li:1024NPE} and $\sigma_{\Delta E}$ (see below).

The NPE takes into account that a potential is defined only up to an additive constant so two curves differing only by a constant should have no error.  This error is defined as 
\begin{equation}
NPE=\max_{i} |E^{\text{FCI}}_{i}-E^{\text{approx}}_{i}|-\min_{i} |E^{\text{FCI}}_{i}-E^{\text{approx}}_{i}|
\end{equation}
where $i$ ranges over all $M$ considered points.  

One possible problem with using the NPE is that two curves with the same maximum and minimum error will have the same NPE regardless of their accuracy for the rest of the points.  We attempt to incorporate the accuracy of the other points by considering the mean squared value of the energy difference. Here the constant $c$, which a potential may be shifted by, is chosen to minimise the sum  
\begin{equation}
S=\frac{1}{M}\sum_{i=1}^{M}\left( \Delta E_{i} -c \right)^{2} 
\end{equation}
where $\Delta E_{i}=E^{\text{FCI}}_{i}-E^{\text{approx}}_{i}$. Setting $\frac{\partial S}{\partial c}=0$ leads to 
\begin{equation}
c=\frac{1}{M}\sum_{i=1}^{M} \Delta E_{i}=\mu_{\Delta E}
\end{equation}
so
\begin{equation}
\min_{c} S=\frac{1}{M}\sum_{i=1}^{M}\left( \Delta E_{i} -\mu_{\Delta E} \right)^{2} =\sigma^{2}_{\Delta E}. 
\end{equation}
This suggests the variance of the difference in energies $\sigma^{2}_{\Delta E}$  as a way to quantify the fit of two potential curves that takes into account all the considered points and that the curves can be shifted by a constant without changing their physics.  To give a quantity in units of energy we then use the standard deviation of $\Delta E$: $\sigma_{\Delta E}$.

\subsection{H\subscript{2}O}

We consider the potential curve for the double hydrogen dissociation of water at a bond angle of $104.5$ degrees with a cc-pVDZ basis and one frozen core.  We generate the FCI results using MOLPRO\cite{MolproFCI1,MolproFCI2,MOLPRO} and use a cut-off of $c_{\text{min}}=10^{-3}$ in the MCCI calculations.  $100$ iterations are used to produce the MCCI natural orbitals here.

\begin{figure}[ht]\centering 
\includegraphics[width=.45\textwidth]{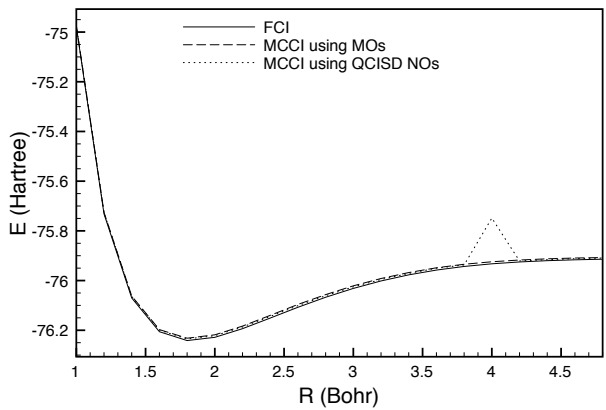}
\caption{Energy (Hartree) against OH bond length $R$ (Bohr) for water in a cc-pVDZ basis with one frozen core using FCI, MCCI ($c_{\text{min}}=10^{-3}$) with either MOs or QCISD NOs.}\label{fig:H2OpotCurve}
\end{figure}

In Fig.~\ref{fig:H2OpotCurve} we see that MCCI with MOs is very close to the FCI curve while when using QCISD NOs in MCCI there is a seemingly anomalous point at $R=4$ Bohr.  This may be linked to the most occupied QCISD natural orbital having an occupation of greater than two  ($2.27$ here) suggesting that the response one matrix is not a good approximation to the actual one matrix.  Non physical occupations of the natural orbitals of the response one matrix of single reference methods has been suggested as a test for when multireference methods are required in Ref.~\onlinecite{GordonNatOrbDiag}.  Interestingly, to four decimal places the largest occupancy is physical for larger $R$ when using QCISD and we can see that the potential curve is again very close to that of the FCI.  The same feature is present when using MP2 natural orbitals.

 Quantifying the accuracy of when using these NOs for the the whole curve would not be useful so we instead first display the results for the fifteen points with $R<4$ Bohr. We include the time necessary for the calculation of the natural orbitals. For all the results, the time for the recalculation of the integrals when using MCCI NOs and the QCISD and MP2 calculation time are all a very small fraction of the total time (less than a second in this case) and are approximately included by using the time for one appropriate geometry ($R=2$ Bohr for the first fifteen points) multiplied by the number of points considered. 
\begin{table}[h]
\centering
\caption{Upper part considering the first fifteen points ($R<4$ Bohr), lower part considering all twenty points for water double hydrogen dissociation in a cc-pVDZ basis. NPE and $\sigma_{\Delta E}$ in kcal/mol.} \label{tbl:H2OFirst15andAll}
\begin{tabular*}{8.5cm}{@{\extracolsep{\fill}}lcccc}
\hline
\hline
Orbitals & NPE & $\sigma_{\Delta E}$ & Mean CSFs & Time (s)  \\
\hline
  MOs  & 3.22  &  0.94 &    1507  & 1200 \\
  MCCI NOs & 4.18 & 1.23 &   1190 & 1491 \\
  MP2 NOs & 1.58 &  0.38 &   1124 & 795  \\
  QCISD NOs & 1.35 &   0.32& 1050 & 694  \\
\hline
  MOs  & 3.22  & 0.92   & 1599     & 1856  \\
  MCCI NOs & 4.49  & 1.14 & 1147 & 2263 \\
  QCISD NOs/MOs & 2.70  & 0.67  &    1256 &  1350     \\
  QCISD NOs/MCCI NOs & 1.35   & 0.37 &     1042 &  1466 \\
\hline
\hline
\end{tabular*}
\end{table}

The upper part of Table \ref{tbl:H2OFirst15andAll} shows that QCISD NOs perform the best over the first fifteen points in terms of accuracy, time and number of CSFs followed by MP2 NOs. It appears that, for bond lengths shorter than $4$ Bohr,  the MCCI NOs perform less well with only the size of the final state smaller than the standard MOs and this is accompanied by a reduced accuracy and longer calculation times.  We suggest that this is because the system is essentially well described by a single reference here and that the $100$ iteration Slater determinant run does not produce natural orbitals,  apart from the largest five, with occupations greater than $0.1$ until $R=3.2$ Bohr. Furthermore the calculation to find the MCCI NOs has a wavefunction consisting of only a single Slater determinant for the smallest two bond lengths.  We note that with only $50$ iterations it was even less likely to produce a MCCI wavefunction consisting of more than one DET which is why we used $100$ iterations for this system.  In this case it would appear that the NPE is less good as we may not do much better compared with using MOs for short bond lengths and may even do worse yet we require more time to calculate the NOs.  However we perhaps do better at larger bond lengths than when using MOs. This possible imbalance may contribute to a larger NPE.

\begin{figure}[ht]\centering 
\includegraphics[width=.45\textwidth]{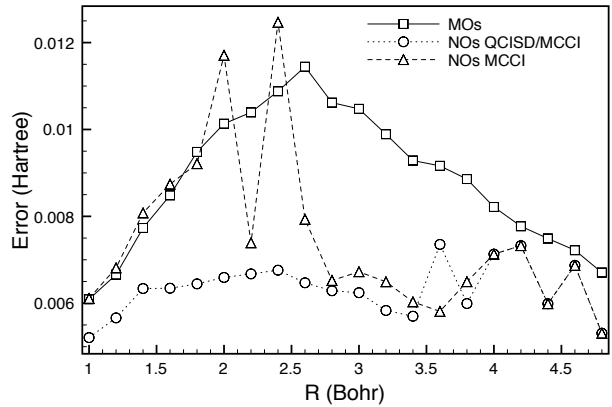}
\caption{Energy error (Hartree) against OH bond length $R$ (Bohr) for water in a cc-pVDZ basis with one frozen core using  MCCI ($c_{\text{min}}=10^{-3}$) with MOs, QCISD/MCCI NOs or MCCI NOs.}\label{fig:H2OpotCurveError}
\end{figure}

We now consider all of the twenty FCI points and use either MOs or MCCI NOs for the last five and do not consider MP2 NOs due to the slightly superior performance of QCISD NOs over the first $15$ points.  We plot the difference between the FCI result and that of MCCI using either molecular or natural orbitals in Fig.~\ref{fig:H2OpotCurveError}.  There we see that the natural orbitals do give similar results to the molecular orbitals when bond lengths are small, but are more accurate at larger bond lengths.  This shows how using QCISD NOs until they are unphysical then proceeding with MCCI NOs results in an accurate curve as the error has a much smaller range than with the other approaches.

  These observations are quantified in the lower part of Table \ref{tbl:H2OFirst15andAll} where it seems to be the case that MCCI NOs do better at longer bond lengths as the NPE is even higher using just MCCI NOs.  The highest accuracy is achieved when using QCISD NOs for points with $R<4$ Bohr then MCCI NOs for larger $R$ where now the NPE is $1.35$ kcal/mol, half the size of when using the same procedure but with MOs instead of MCCI NOs.  The time required is slightly longer when using the MCCI NOs but represents an increase of less than ten percent compared with the NPE halving.  The mean number of CSFs at $1042$ is a substantial reduction of the FCI space which consists of around $8\times 10^{7}$ Slater determinants when spatial symmetries are neglected.  

This suggests the approach of using QCISD natural orbitals until the natural orbital occupations become unphysical then switching to MCCI natural orbitals.  This should mean that a good approximation to the natural orbitals is achieved by QCISD when the correlation is essentially dynamic and then by MCCI when static correlation becomes important.

The results show that $\sigma_{\Delta E}$ and NPE have the same behaviour with one slight difference being that QCISD NOs over 15 points and QCISD NOs/ MCCI NOs over 20 points have the same NPE to two decimal places but  $\sigma_{\Delta E}$ increases a little for QCISD NOs/ MCCI NOs showing a small decrease in accuracy that is not revealed with the NPE. Using MCCI NOs for the last five points halves the NPE compared with using MOs, but the $\sigma_{\Delta E}$ value is $0.55$ of its previous value.

\subsubsection{aug-cc-pVTZ}
\begin{figure}[ht]\centering 
\includegraphics[width=.45\textwidth]{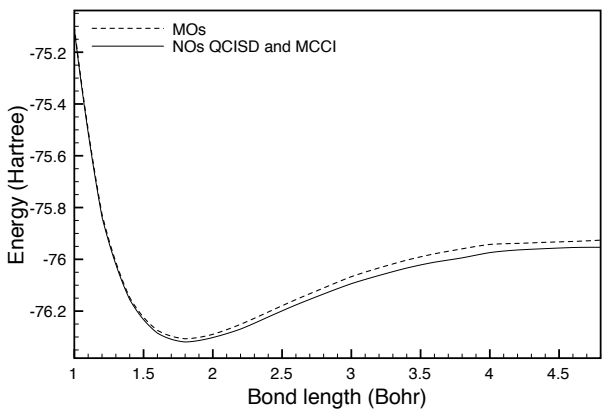}
\caption{Energy (Hartree) against  bond length (Bohr) of both hydrogens for water with an aug-cc-pVTZ basis using  MCCI ($c_{\text{min}}=10^{-3}$) with either MOs or NOs.}\label{fig:H2OpotCurveAugVTZ}
\end{figure}

We now increase the basis size to aug-cc-pVTZ while keeping other parameters the same.  Fig.~\ref{fig:H2OpotCurveAugVTZ} shows that the potential curves behave generally as expected and when using natural orbitals the energy is noticeably lower as dissociation is approached.  There are no FCI results for comparison but the curves and results for the cc-pVDZ basis suggest that the NO method should be more accurate.

For the first fifteen points where the QCISD response one matrix is physical the calculation is also substantially faster:  $0.98$ hours versus 4.6 hours.  Furthermore the wavefunctions require fewer CSFs with an average of 1681 CSFS when using QCISD natural orbitals compared with 5744 CSFs when using MOs for the first fifteen points. When using MCCI NOs for the longer bond lengths and considering all points we see in Table \ref{tbl:H2OaugVTZallPoints} that the calculation is substantially faster, although the improvement is not as great as that seen over the first fifteen points, and the mean number of CSFs is also much smaller.

\begin{table}[h]
\centering
\caption{Results for all points for the double hydrogen dissociation of water in an aug-cc-pVTZ basis.} \label{tbl:H2OaugVTZallPoints}
\begin{tabular*}{8.5cm}{@{\extracolsep{\fill}}lcc}
\hline
\hline
Orbitals  & Mean CSFs & Time (Hours)  \\
\hline
  MOs   & 5825     &  7.15 \\
  QCISD NOs/MCCI NOs  & 1924 &  2.77      \\
\hline
\hline
\end{tabular*}
\end{table}

We note that, when neglecting symmetry, the number of DETs in the FCI space has increased from $8\times 10^{7}$ to around $7\times 10^{12}$ when using this larger basis, which is approximately $9\times 10^{4}$ as many DETs. However MCCI with QCISD NOs takes $3.4$ times as long and with $2.4$ times as many CSFs for the points which it can be applied to compared with results for the method using cc-pVDZ.  For all the points, using MOs gave a time scaling of $13.9$ and a scaling of $3.6$ for CSFs compared with the cc-pVDZ MCCI MO calculations.  When using QCISD NOs/MCCI NOs the time scaling was around $6.8$ and about $1.9$ times as many CSFs were required compared with this method using a cc-pVDZ basis.  
The scalings appear promising when compared with the growth in the size of the FCI space and can hopefully be further improved.

\subsubsection{Excited state}

\begin{figure}[ht]\centering 
\includegraphics[width=.45\textwidth]{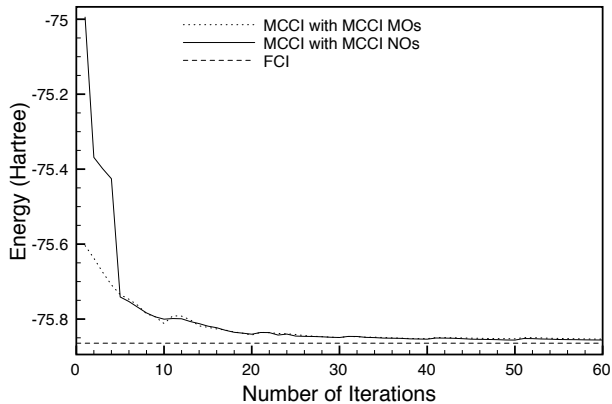}
\caption{Energy (Hartree) against number of iterations for water in a cc-pVDZ basis with no frozen cores and an OH bond length of $R=2$ Bohr using  MCCI ($c_{\text{min}}=10^{-3}$) with MOs or MCCI NOs compared with the FCI result.}\label{fig:H2Oexcite}
\end{figure}

We briefly return to water in a cc-pVDZ basis and as an aside we demonstrate the use of MCCI natural orbitals for an excited state.  Here the other types of approximate natural orbitals considered are not available.  We note that the current version of MCCI calculates one eigenvalue with the Davidson algorithm so the diagonalization routine can become unstable when dealing with excited states: the program may find itself in a subset of the CSF space so that the previous excited state of interest is now the ground state for example.  Hence we only consider one geometry: the first excited state of $A_{1}$ symmetry for water in $C_{2v}$ with $R=2$ Bohr.  We now only use fifty iterations to create the MCCI NOs as many DETs are found for the first excited state. Furthermore no orbitals are frozen, however we still employ the approximate NOs in an MCCI CSF calculation. We see in Fig.~\ref{fig:H2Oexcite} that when using the MCCI NOs the energy is initially substantially higher than with MOs but rapidly decreases and becomes slightly lower than that due to MCCI MOs at convergence.  The final MCCI wavefunction uses fewer CSFs when NOs are employed here:  2308 versus 1755.  With MOs the time to convergence was 149 seconds while only 102 seconds were needed using MCCI NOs however the calculation of the NOs was more involved here so the total time when using MCCI NOs was around 244 seconds.  It would appear that fewer iterations for the calculation of the MCCI NOs may be useful for reducing the total calculation time.

  Future work is planned using state-averaging of the MCCI wavefunction for a few states to reduce instabilities in the calculation and enable better calculation of excited potential energy curves.  There is also the possibility of other spin states being reached in the MCCI DET calculation and the consideration of more than one eigenvalue may allow better discrimination of these spin states.  The reasonably promising result for the use of NOs in the calculation of an excited state should hopefully be improved upon using these approaches.   
\subsection{N\subscript{2}}

We now consider the MCCI potential energy curve for N\subscript{2} dissociation with two frozen cores in a cc-pVDZ basis.  Here fifty iterations are used to create the MCCI NOs. The fifteen FCI results are gathered from Refs.~\onlinecite{LarsenN2FCI2000,GwaltneyN2FCI2002,chanN2FCI2002}.

Similar to our findings for water the MCCI run for small $R$ does not result in a state beyond that comprising the occupied MOs.  This occurs for both cut-offs we consider, so we continue with the use of QCISD NOs until they become unphysical when we switch to MCCI NOs.  This does not occur until the last FCI point (2.225 angstroms) in this case.  We see in Table \ref{tbl:N2nos} that the use of approximate natural orbitals reduces the calculation time and the average number of states required. The accuracy is also improved by the use of natural orbitals here.

\begin{table}[h]
\centering
\caption{N\subscript{2} results with cc-pVDZ and $c_{\text{min}}=10^{-3}$. NPE and $\sigma_{\Delta E}$ in kcal/mol.} \label{tbl:N2nos}
\begin{tabular*}{8.5cm}{@{\extracolsep{\fill}}lccccc}
\hline
\hline
Orbitals & NPE & $\sigma_{\Delta E}$ & Mean CSFs & Time (Hours)  \\
\hline
  MOs  & 6.37 & 1.69   &  2909     & 1.69   \\
  QCISD NOs/MCCI NOs  & 5.03  & 1.39   & 2478   & 1.10       \\ 
\hline
\hline
\end{tabular*}
\end{table}

With a smaller cut-off, Table \ref{tbl:N2nossmallercmin} shows that the calculation takes longer but the speedup due to the use of approximate natural orbitals is of a similar factor.  The improvement in accuracy is not quite such a large scaling as for the smaller cut-off but again the approximate NOs have improved calculation time and accuracy.
\begin{table}[h]
\centering
\caption{N\subscript{2} results with cc-pVDZ and $c_{\text{min}}=5\times10^{-4}$. NPE and $\sigma_{\Delta E}$ in kcal/mol.} \label{tbl:N2nossmallercmin}
\begin{tabular*}{8.5cm}{@{\extracolsep{\fill}}lccccc}
\hline
\hline
Orbitals &  NPE & $\sigma_{\Delta E}$ & Mean CSFs & Time (Hours)  \\
\hline
   MOs  &   3.98   &  1.06     & 7185   & 7.55 \\
  QCISD NOs/MCCI NOs &   3.49  & 0.87    & 5758   & 4.98       \\   
\hline
\hline
\end{tabular*}
\end{table}

\begin{figure}[ht]\centering 
\includegraphics[width=.45\textwidth]{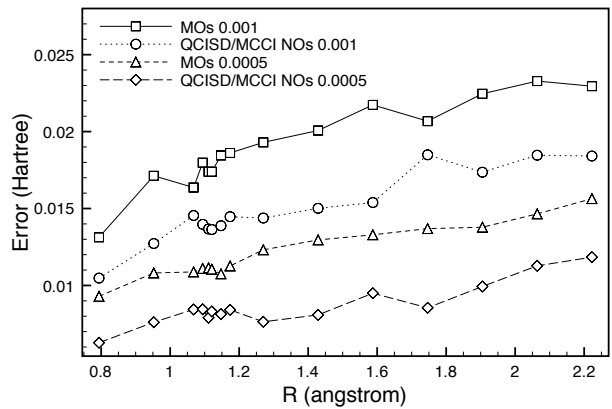}
\caption{Energy error (Hartree) against  bond length (Bohr) for N\subscript{2} when using MCCI with a cc-pVDZ basis, two different cut-off values and either MOs or approximate NOs. }\label{fig:N2natpotError}
\end{figure}

We see in Fig.~\ref{fig:N2natpotError} that the error of the MCCI results when compared with FCI decreases when approximate natural orbitals are used and when the cut-off is lowered from $c_{\text{min}}=10^{-3}$ to $c_{\text{min}}=5\times10^{-4}$.  The reduction due to the smaller cut-off is greater than that due to using approximate NOs.

\subsection{CO}

We use a cc-pVDZ basis set to model the dissociation of carbon monoxide and freeze two of the orbitals.  For MCCI the cut-off is $c_{\text{min}}=5\times 10^{-4}$ and $50$ iterations are used for the generation of the MCCI NOs. The FCI space consists of $4\times10^{9}$ Slater determinants when neglecting symmetry. We find that the QCISD NOs become unphysical at $R=3.2$ Bohr here and if we consider the $17$ points for bond length smaller than this we see in Table \ref{tbl:COfirst17Points} that accuracy, time and size of the wavefunction are all improved by the use of NOs.  Interestingly the most accurate curve for the first seventeen points is due to the MCCI NOs in contrast to the results for water and N\subscript{2}. Now the MCCI occupied NOs are never just the occupied MOs.  The fastest calculation and fewest CSFs on average both belong to the calculation using QCISD NOs.   
 
\begin{table}[h]
\centering

\caption{Results for the first $17$ points for CO ($R\leq 3$ Bohr). NPE and $\sigma_{\Delta E}$ in kcal/mol.} \label{tbl:COfirst17Points}
\begin{tabular*}{8.5cm}{@{\extracolsep{\fill}}lcccc}
\hline
\hline
Orbitals & NPE & $\sigma_{\Delta E}$ & Mean CSFs & Time (Hours)  \\
\hline
  MOs  & 3.11  & 0.89   & 7053     & 6.76 \\
  QCISD NOs & 2.77  & 0.80  &  5616 &  4.74     \\
    MCCI NOs & 1.58   & 0.41  &    6069 &  5.79 \\
\hline
\hline
\end{tabular*}
\end{table}

The MP2 natural orbitals become unphysical sooner than those of QCISD for this system: the largest occupation is around $2.03$ at $3$ Bohr but some small negative higher occupations occur at shorter bond lengths.
The MCCI point at $3$ Bohr is then of poor accuracy.  If we exclude this and compare over the first 16 points we have a NPE of 8.28 kcal/mol compared with 2.75 kcal/mol when using the QCISD natural orbitals.  This poor performance seems to be due to the occurrence of negative, although small, natural orbital occupations.

\begin{figure}[ht]\centering 
\includegraphics[width=.45\textwidth]{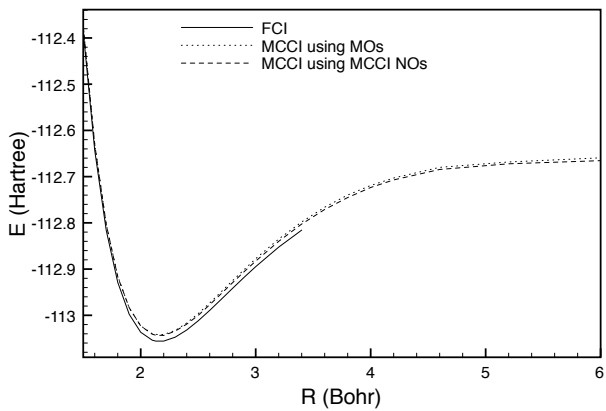}
\caption{Energy (Hartree) against  bond length $R$ (Bohr) for CO with a cc-pVDZ basis using FCI, MCCI ($c_{\text{min}}=5\times 10^{-4}$) with either MOs or QCISD NOs.}\label{fig:COpotCurve}
\end{figure}

We see in Fig.~\ref{fig:COpotCurve} that the curves appear to be well behaved and are close to the FCI points where they are available.  We note that we were unable to calculate FCI points for larger bond lengths due to disk space requirements.  For the $19$ points for which we have FCI results the NPE values in kcal/mol are $1.58$ for MCCI NOs, $3.37$ for MOs, $3.40$ for QCISD NOs/MCCI NOs while the $\sigma_{\Delta E}$ values in kcal/mol are respectively  $0.39$, $1.01$ and  $0.91$.  It is interesting that the order of MOs and QCISD NOs/MCCI NOs with regards to accuracy changes in this case depending on whether it is quantified using the NPE or $\sigma_{\Delta E}$, although the differences are small.
 
\begin{figure}[ht]\centering 
\includegraphics[width=.45\textwidth]{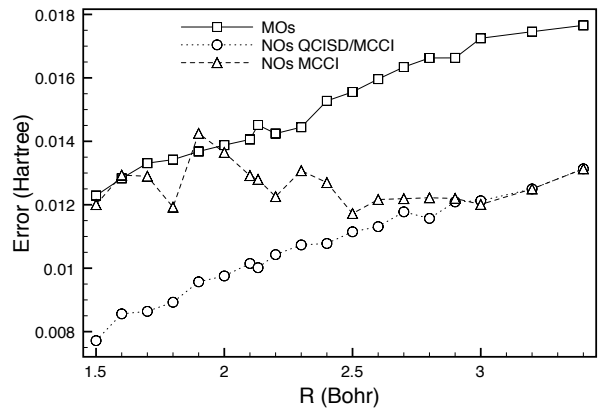}
\caption{Energy error (Hartree) against  bond length $R$ (Bohr) for CO with a cc-pVDZ basis for MCCI ($c_{\text{min}}=5\times 10^{-4}$) with either MOs, QCISD/MCCI NOs or MCCI NOs.}\label{fig:COpotCurveError}
\end{figure}

The energy error when compared with the available FCI results is depicted in Fig.~\ref{fig:COpotCurveError}.  This reveals that the lowest error is achieved when using QCISD NOs however the error increases with bond length until it becomes similar to that found when using MCCI NOs.  The smallest range of errors comes from using MCCI NOs which results in this approach having the lowest NPE and $\sigma_{\Delta E}$.

We see in Table \ref{tbl:COallPoints} that, when all points forming the curve are considered, the use of approximate natural orbitals accelerates the calculation, uses fewer CSFs and the potential energy curves suggest that there should be an improvement in accuracy.  The fastest was a combination of QCISD and MCCI NOs but the results for the first $19$ points suggest that the most accurate results in this case would perhaps be due to MCCI NOs.

\begin{table}[h]
\centering
\caption{Results considering all $26$ points for CO.} \label{tbl:COallPoints}
\begin{tabular*}{8.5cm}{@{\extracolsep{\fill}}lcc}
\hline
\hline
Orbitals & Mean CSFs & Time (Hours)  \\
\hline
  MOs  &  8,208     & 20.72   \\
  QCISD NOs /MCCI NOs &  6,993 &  17.88    \\
  MCCI NOs & 7,289 &  18.93 \\
\hline
\hline
\end{tabular*}
\end{table}

We note that the $17$ FCI points up to and including $R=3$ Bohr required around $709$ processor hours which we could approximately equate to 59 Hours when running on 12 processors.  This is still ten times slower than the MCCI  calculation using MCCI NOs, over the same number of points and furthermore storage space issues meant that the FCI calculation could not be run to convergence for $R\geq 3.6$.

\section{Second-order perturbation theory}

The second-order perturbation scheme for configuration interaction in Ref.~\onlinecite{HarrisonFCIperturbation} considers an energy lowering
\begin{equation}
\nonumber \Delta E_{K}=\sum_{I} \frac{|\bra{I} \hat{H} \ket{K} |^{2}}{E_{K}-\bra{I} \hat{H} \ket{I}}.
\end{equation}
Here $\ket{K}$ is the current CI wavefunction while the sum is over all $\ket{I}$ which are formed by single and double substitutions from $\ket{K}$.  If a contribution from any $\ket{I}$ is greater than a threshold then these $\ket{I}$ are added to the reference space and a new wavefunction found by diagonalising the Hamiltonian. The process is continued until no new states are added to the CI wavefunction and then the final  $\Delta E_{K}$ gives an estimate of the energy lowering due to the neglected states.  We use this scheme with the final wavefunction from a MCCI calculation to attempt to account for more of the dynamic correlation (MCCIPT2).  For this we use an MCCI version where states are again added on a step following one where all states have been considered for removal.  We note that the program is run for $200$ iterations on eight processors without a convergence check here.

If we write $\hat{H}=\hat{H_0}+\hat{H}'$ and have $\hat{H}_0 \ket{\Psi_{MCCI}}=E_{MCCI}\ket{\Psi_{MCCI}}$. Then for Slater determinants we have $\bra{I} \hat{H} \ket{K} = \bra{I} \hat{H}' \ket{K}$, but for the non-orthogonal CSFs used in MCCI we need to use $\bra{I} \hat{H} \ket{K}-E_{K}\bra{I} K \rangle = \bra{I} \hat{H}' \ket{K}$ in the numerator to give.

\begin{equation}
\nonumber \Delta E_{K}=\sum_{I} \frac{|\bra{I} \hat{H} \ket{K}-E_{K}\bra{I} K \rangle  |^{2}}{E_{K}-\bra{I} \hat{H} \ket{I}}.
\end{equation}
Here all states are normalised. We use $c_{\text{min}}$ as the threshold to consider if a state, in the PT2 scheme, should be added to the MCCI wavefunction.  We note that for CSFs we use the same procedure\cite{mcciGreer98} of a random walk through the branching diagram as in MCCI.  This followed by the removal of duplicates ensures that the CSFs are linearly independent but may mean that it is conceivable that some CSFs are neglected.

The slowest step in the original PT2 method was checking if a prospective state was a duplicate in the set of all single and double substitutions or if it should be added to them.\cite{HarrisonFCIperturbation}  Given the size of $I$ as $N_{I}$ then this requires $O(N_{I}^{2})$ operations if we check each new member against all previous and assume that the size of the space without duplicates is approximately a constant fraction of the size of $I$.  As $N_{I}$ is expected to be very large compared with the states in the MCCI wavefunction, we consider sorting the list of $I$ by alpha and beta string using the quicksort algorithm.\cite{Hoare62} This will tend to need $O(N_{I}\log(N_{I}))$ operations followed by one pass through the sorted list of $O(N_{I})$ to delete repeated states. We also have to delete any members of $K$ in $I$ but this is quick as $K$ is small in comparison with $I$.  The set of $I$ can then be split amongst processors to calculate $\Delta{E}$ in parallel but this is currently implemented only in the case of Slater determinants. We note that a small test calculation with  $10$ CSFs in the final MCCI wavefunction required ten times longer when not using the new method of removing duplicates. 

We test MCCIPT2 on N\subscript{2} in a cc-pVDZ basis with two frozen cores and a MCCI cut-off of $c_{\text{min}}=10^{-3}$. The MCCI calculations are carried out using eight processors. Two hundred iterations are used for each MCCI calculation.

Slater determinants did not work so efficiently here: new states were discovered when using MCCIPT2 and we found that running another MCCI calculation each time with the reference taken as the last MCCI wavefunction plus the added PT2 states until no new states were found was necessary to achieve a smooth potential curve.  Nevertheless the use of MCCIPT2 improved the accuracy from an NPE of 11.01 kcal/mol for MCCI with PT2 states to an NPE for 6.53 kcal/mol for MCCIPT2 while $\sigma_{\Delta E}$ reduced from 3.12 kcal/mol to 1.86 kcal/mol.

\begin{figure}[ht]\centering 
\includegraphics[width=.45\textwidth]{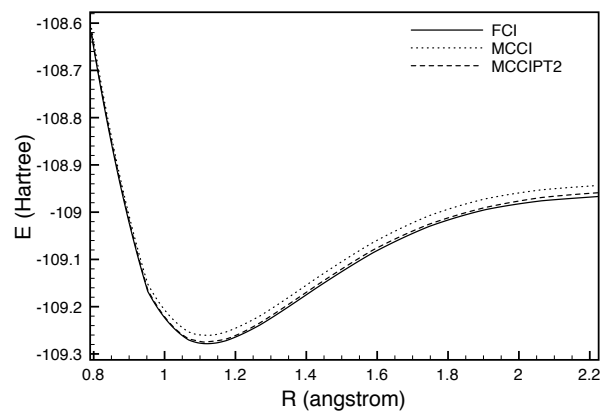}
\caption{Energy (Hartree) against  bond length (angstrom) for N\subscript{2} with a cc-pVDZ basis using MCCI and MCCIPT2 with CSFs and $c_{\text{min}}=10^{-3}$, compared with FCI results.\cite{LarsenN2FCI2000,GwaltneyN2FCI2002,chanN2FCI2002}}\label{fig:MCCIPT2csfN2}
\end{figure}

\begin{figure}[ht]\centering 
\includegraphics[width=.45\textwidth]{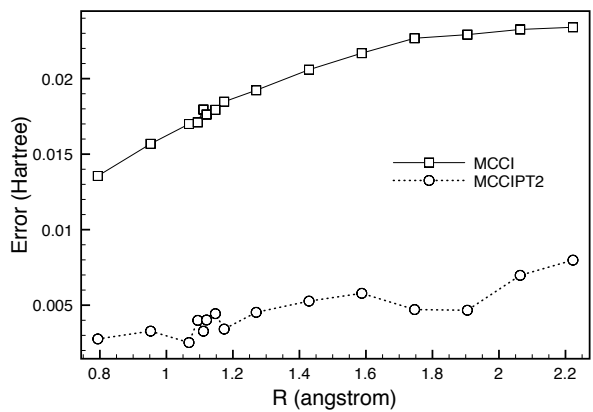}
\caption{Energy error (Hartree) against  bond length (angstrom) for N\subscript{2} with a cc-pVDZ basis when using MCCI or MCCIPT2 with CSFs and $c_{\text{min}}=10^{-3}$. }\label{fig:MCCIPT2csfN2Error}
\end{figure}

When using CSFs no states were found by the PT2 procedure with a large enough contribution to be added to the MCCI wavefunction.  This suggests that, with regards to our requirement for adding states using PT2, the MCCI wavefunction is, in a sense, optimum when using CSFs here.   We see in Fig.~\ref{fig:MCCIPT2csfN2} that the MCCIPT2 curve appears to be of higher accuracy as at times it is difficult to distinguish from the FCI curve on the scale of the graph. The plot of differences between the MCCI and FCI energies (Fig.~\ref{fig:MCCIPT2csfN2Error}) shows that the errors are much smaller using MCCIPT2 and a little more balanced.  The NPE for MCCI here was similar to previous MCCI calculations for nitrogen at $6.18$ kcal/mol and this was reduced to 3.42 kcal/mol when using MCCIPT2.  The  $\sigma_{\Delta E}$ value was lowered from 1.83 kcal/mol to 0.92 kcal/mol by using MCCIPT2.

The time for a a single-point MCCI calculation on 8 processors of $200$ iterations ranged from less than one minute to around $1.3$ hours as $R$ increased here.  While the total time including the PT2 calculation on one processor for this proof of concept program ranged from less than four minutes to almost 2 hours as R increased. The number of states comprising the MCCI wavefunction ranged from around $1000$ to almost $5000$ as $R$ increased while the states in the PT2 energy lowering calculation ranged from $0.4$ million to $1.6$ million.  The accuracy of MCCIPT2 at $c_{\text{min}}=10^{-3}$ was better than MCCI at $c_{\text{min}}=5\times 10^{-4}$ (see Table \ref{tbl:N2nossmallercmin}) however the time was longer at around $10.7$ hours but this was on $8$ processors and without a convergence check. If we consider the time for PT2 only (4.18 Hrs) and reasonably assume it would be similar if used on the MO MCCI $c_{\text{min}}=10^{-3}$ results of Table \ref{tbl:N2nos} then this suggests a time of around 5.9 Hours which would be faster than MCCI with MOs at $c_{\text{min}}=5\times 10^{-4}$ but not MCCI with approximate natural orbitals at this cut-off.     The results are encouraging and the PT2 CSF code for MCCI has room for improvement, e.g., parallelisation and more efficient calculation of matrix elements.

\section{Summary}

We introduced a way to approximate natural orbitals in MCCI and we have seen that approximate natural orbitals from an MP2, QCISD or MCCI run could reduce the time and number of states necessary for a single-point MCCI calculation when using Slater determinants. We introduced a measure of accuracy of a potential curve ($\sigma_{\Delta E}$) that takes into account that the curve can be shifted by a constant but, unlike the non-parallelity error, considers all points in the curve.  For the curves considered in this paper the behavior of each measure was usually similar although there were occasions when the NPE did not change but $\sigma_{\Delta E}$ did, or that the ordering of accuracy using the two approaches was changed for small differences.

For the potential curve for double hydrogen dissociation of water in a cc-pVDZ basis we found that if the QCISD or MP2 natural orbitals became unphysical the accuracy of the MCCI potential curve could be severely impacted.  The results suggested the use of QCISD natural orbitals until they had occupations greater than two or negative occupations, then switching to MCCI natural orbitals (QCISD NOs/MCCI NOs) offered the largest improvement in accuracy and number of CSFs and took less time than when using molecular orbitals.  Similar results were seen for the potential curve for N\subscript{2} in a cc-pVDZ basis.  We noted that the MCCI natural orbitals could be unsuitable at bond lengths where a single reference was a good approximation as here the only occupied MCCI natural orbitals were the occupied molecular orbitals when using short MCCI calculations.  We used the approach of QCISD NOs/MCCI NOs for a potential curve of water in an aug-cc-pVTZ basis and saw good improvements in calculation time and the number of CSFs required.  The scaling in calculation time compared with the cc-pVDZ basis was very much smaller than the increase in the size of the FCI space.

For the potential curve for the dissociation of carbon monoxide the MCCI potential curve was most accurate when using MCCI natural orbitals for the points that we had FCI results for.  The calculation time for the entire curve was a little longer than when using QCISD then MCCI natural orbitals but was still better than using molecular orbitals.  We note that the use of approximate natural orbitals here did not always improve convergence or reduce the error. However by using QCISD NOs/MCCI NOs in MCCI calculation speed and accuracy were seen to be increased when compared with results using MOs.  This small sample of molecules seems to suggest that the MCCI natural orbitals should be used unless there are many MCCI natural orbitals with zero occupation at the start of the curve then QCISD NOs/MCCI NOs should be employed.

We saw that an adaptation of a second-order perturbation scheme\cite{HarrisonFCIperturbation} combined with MCCI (MCCIPT2) could run faster when using a new method to remove duplicates in the space of single and double substitutions of the reference.  We found that at the same level of cut-off, the MCCIPT2 calculation with Slater determinants was much less efficient than that with CSFs.  MCCIPT2 gave results with higher accuracy than the MCCI calculation alone for the potential curve of the dissociation of the nitrogen molecule.

\acknowledgements{We thank the European Research Council (ERC) for funding under the European Union's Seventh Framework Programme (FP7/2007-2013)/ERC Grant No. 258990.}

\providecommand{\noopsort}[1]{}\providecommand{\singleletter}[1]{#1}%
\end{document}